\documentclass[conference]{IEEEtran}
\IEEEoverridecommandlockouts

\usepackage{cite}
\usepackage{amsmath,amssymb,amsfonts}
\usepackage{algorithmic}
\usepackage{graphicx}
\usepackage{booktabs}
\usepackage{microtype}
\usepackage{etoolbox,siunitx}
\robustify\bfseries
\usepackage{textcomp}
\usepackage{xcolor}
\def\BibTeX{{\rm B\kern-.05em{\sc i\kern-.025em b}\kern-.08em
    T\kern-.1667em\lower.7ex\hbox{E}\kern-.125emX}}
\begin{document}

\title{Data Augmentation Using Neural Acoustic Fields With Retrieval-Augmented Pre-training
\thanks{This work was performed while C.~Ick was an intern at MERL.}
}

\author{
    \IEEEauthorblockN{\textit{
        Christopher Ick\IEEEauthorrefmark{1}\IEEEauthorrefmark{2},
        Gordon Wichern\IEEEauthorrefmark{1}, %
        Yoshiki Masuyama\IEEEauthorrefmark{1}, %
        François G.\ Germain\IEEEauthorrefmark{1}, %
        and Jonathan Le Roux\IEEEauthorrefmark{1}}
        \vspace{.7\baselineskip}}
        \IEEEauthorblockA{
        \IEEEauthorrefmark{1}Mitsubishi Electric Research Laboratories (MERL), Cambridge, MA, USA}
        \IEEEauthorblockA{
        \IEEEauthorrefmark{2}Music and Audio Research Laboratory, New York University, Brooklyn, NY, USA}
}

\maketitle

\begin{abstract}
This report details MERL's system for room impulse response (RIR) estimation submitted to the Generative Data Augmentation Workshop at ICASSP 2025 for Augmenting RIR Data (Task 1) and Improving Speaker Distance Estimation (Task 2).
We first pre-train a neural acoustic field conditioned by room geometry on an external large-scale dataset in which pairs of RIRs and the geometries are provided.
The neural acoustic field is then adapted to each target room by using the enrollment data, where we leverage either the provided room geometries or geometries retrieved from the external dataset, depending on availability.
Lastly, we predict the RIRs for each pair of source and receiver locations specified by Task 1, and use these RIRs to train the speaker distance estimation model in Task 2.
\end{abstract}


\section{Introduction}
The goal of the Room Acoustics and Speaker Distance Estimation Challenge, which is part of the Generative Data Augmentation Workshop at ICASSP 2025, is to design systems for room impulse response (RIR) estimation at unseen locations from a small set of training examples~\cite{GenDARA}. The challenge is split into two tasks, Augmenting RIR Data (Task 1), and Improving Speaker Distance Estimation (Task 2).

Interest in RIR estimation has surged in recent years, with the successful application of room acoustics estimation in immersive experiences for virtual/extended reality, adaptive sound reproduction systems, acoustic scene understanding, and more~\cite{tewary2021advances}.
Typical settings require a system to predict an RIR from an existing set of acoustic measurements.
With modern approaches utilizing deep neural networks, the use of generated acoustic data for augmentation has been essential for training machine learning-based systems.

In our approach for Task 1, we use neural acoustic fields to build spatially continuous representations of acoustic spaces.
We use an acoustic similarity-based retrieval method to build a room-specific pre-training dataset, and then fine-tune our model on limited training examples using low-rank adaptation (LoRA) ~\cite{hu2022lora}.
After pre-training and fine-tuning, the model can be used to estimate the RIR at unseen source/receiver locations within the specified room.

\section{Preliminaries}
In Task 1, we are asked to generate RIRs for twenty different rooms.
For each room, between 5 and 10 RIRs at different source and receiver locations are provided in the enrollment set.
For rooms 1--10, room geometry information is provided in the form of a room mesh object file.
For rooms 11--20, only the RIRs and source/listener locations are provided.
An additional room (Room 0) is provided with recorded and simulated RIRs at 20 source/listener locations, and an additional 405 simulated RIRs at grid locations within the room.
The goal for Task 1 is to design a system to estimate the RIRs in each of the twenty rooms at unseen source/receiver locations specified by the organizers.
The generated RIRs are then evaluated on a variety of acoustic quality measures, namely RT60, EDF, and DRR~\cite{GenDARA}.

\section{Methodology}
\subsection{Retrieval for Pre-training}

\begin{figure}
    \centering
    \includegraphics[width=\linewidth]{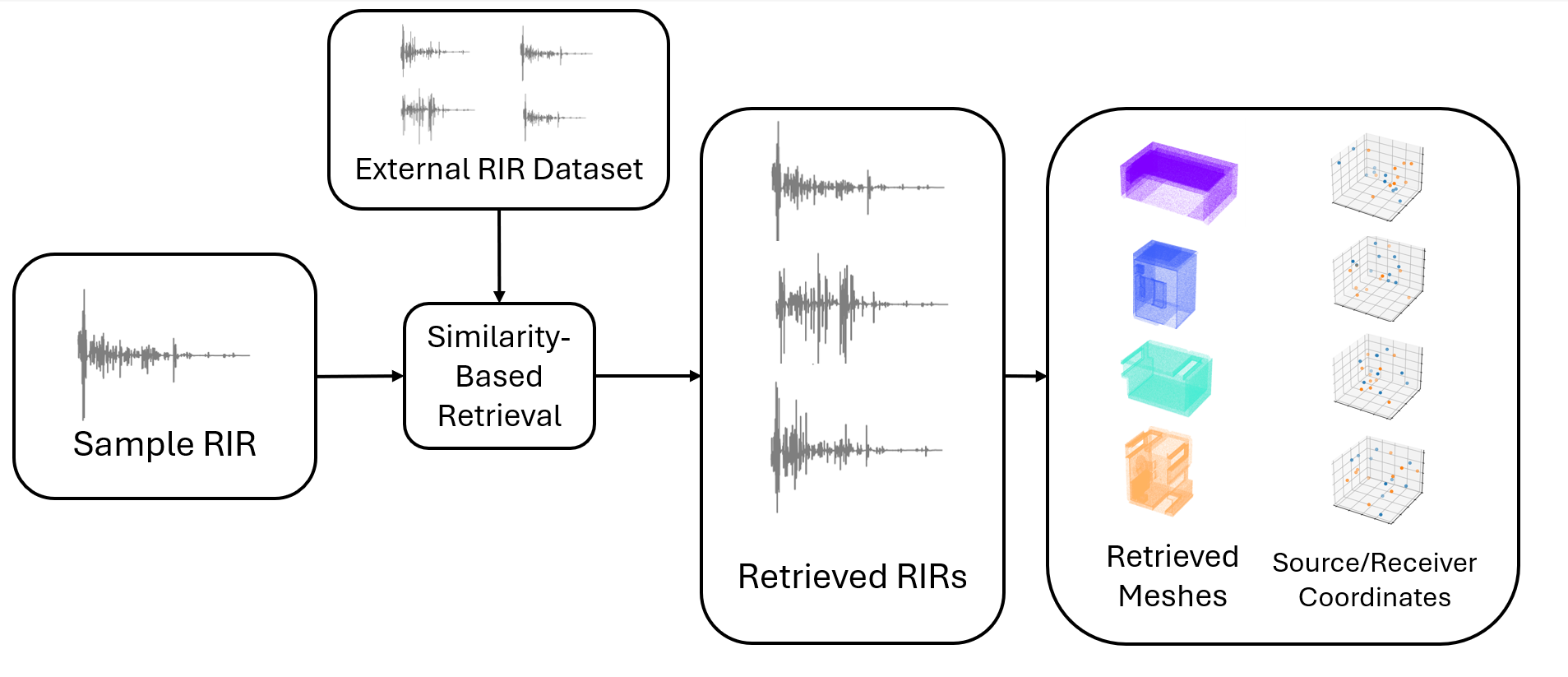}
    \vspace{-.75cm}
    \caption{Retrieval strategy for pre-training dataset selection from a sample RIR.}
    \label{fig:retrieval}
\end{figure}

Each room $\{R_i\}_{i=1}^{20}$ is provided with $N$ RIRs $\{h_{i,j}\}_{j=1}^{N}$, and pairs of corresponding source and receiver positions $(s_{i,j}, r_{i,j})$.
To find appropriate data for pre-training, we use each room's provided RIRs to retrieve RIRs from an external dataset.
We use the GWA dataset \cite{tang2022gwa}, a large-scale RIR dataset that contains over 2 million RIRs that were simulated based on 3D scene meshes from the 3D-FRONT dataset  \cite{fu20213dls,fu20213dst} to ensure acoustic diversity.
Each scene in the 3D-FRONT dataset contains multiple rooms and objects; by building bounding boxes defined by the boundaries of the scene meshes, we defined a set of rooms $\{\tilde{R}_i\}_{i>20}$ corresponding to different regions within the 3D-front scenes.

To select an appropriate set of rooms from this dataset, for each room, we use a similarity-based retrieval method shown in~\ref{fig:retrieval}.
For each RIR $h_{i,j}$ in the enrollment data for room $R_i$, we compute a multi-band RT60 with $B$ bands $\text{RT}_{60}(h_{i,j})\in\mathbb{R}_+^B$.
With it, we can query the dataset by using the $L2$ distance between each measured multi-band RT60 and that of each of the RIR samples $\tilde{h}_l$ in the GWA set $\|\text{RT}_{60}(h_{i,j})-\text{RT}_{60}(\tilde{h}_{l})\|_2$, selecting $M$ closest RIRs in the dataset for each of the $N$ enrollment RIRs.
For each of these $N \times M$ retrieved RIRs, we collect the room(s) containing the source and receiver.
We can then rank these rooms based on their frequency in the set of retrieved RIRs, providing us with a distribution of matching rooms, each with several RIR measurements at various source/receiver locations.
In our experiments, for training efficiency, we limited these rooms to the 100 most frequently retrieved rooms.
This set of rooms can be used to create a room-specific pre-training set for each of our 20 rooms.
We compare this retrieval method with randomly selected rooms in \ref{tab:results}.
In the cases of rooms 11--20, where no room mesh is provided, we further retain the most closely matched room's geometry for use as the estimated geometry for the queried room.

\subsection{Neural Acoustic Field}
For our model, we use a simplified single-channel version of the INRAS architecture \cite{su22inras}, excluding orientation information.
This model takes a pair of source and listener positions in 3D space, as well as geometric features in the form of sampled points from the surface of the room mesh (referred to as bounce points).
Following \cite{chen2024RAF}, for each room, we evenly sample the room's mesh using Poisson disk sampling, such that we are provided with $K$ evenly spaced points characterizing the room's geometry.
These bounce points, as well as the coordinates of the source and receiver, are used as inputs to our model.
The model then uses a sinusoidal encoding and a multilayer perceptron to generate representations of the room's IR measurement at a given pair of source and receiver locations.
After training, this latent representation can then be used to generate RIRs at an unseen pair of source and receiver positions.

\begin{figure}
    \centering
    \includegraphics[width=0.8\linewidth]{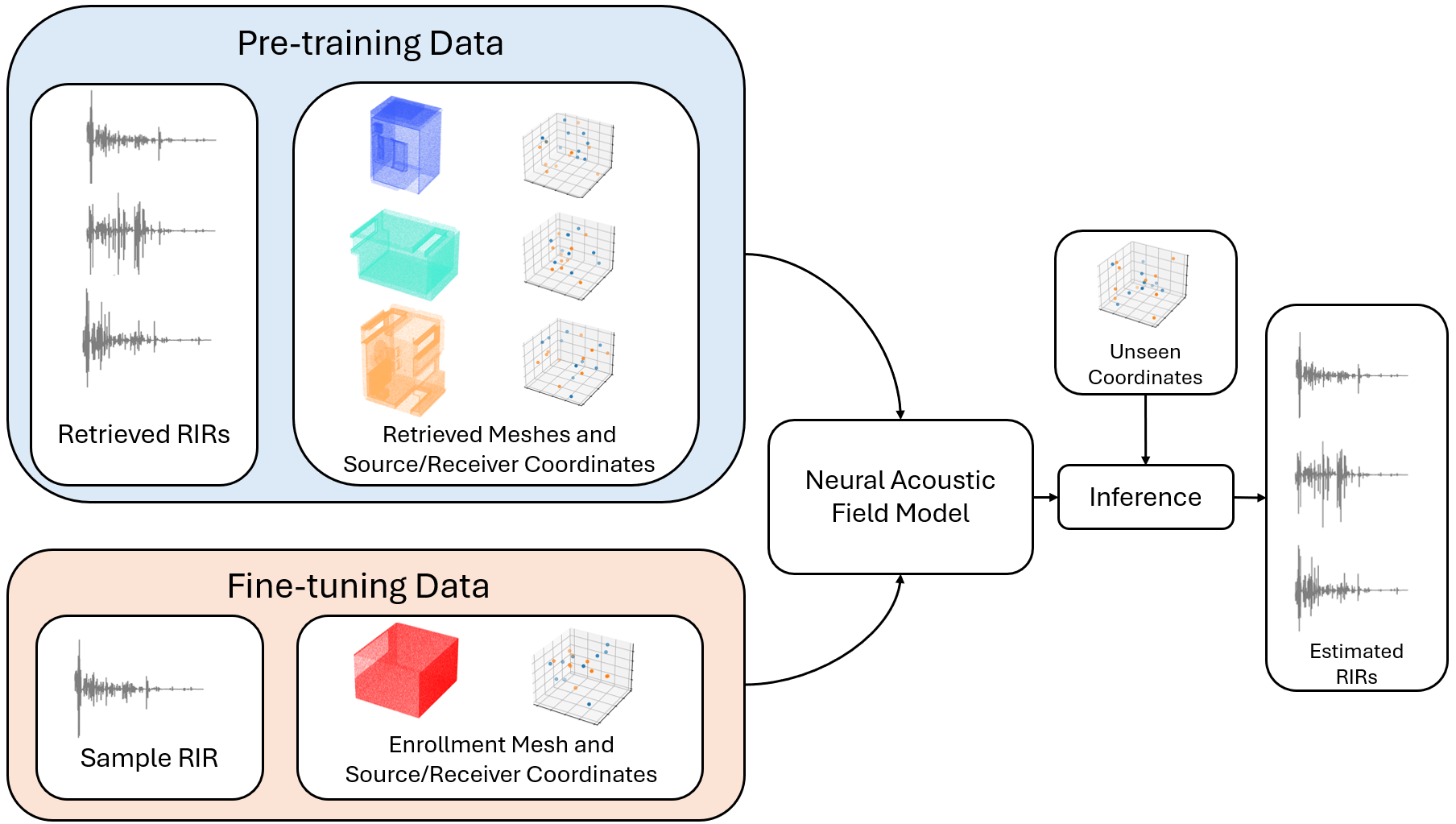}
    \caption{Pre-training and fine-tuning strategies for neural acoustic field training.}
    \label{fig:training}
\end{figure}

\subsection{Fine-Tuning}
During pre-training, we expose the model to a wide range of source and receiver location pairs inside a variety of room geometries.
After this step, we can then fine-tune the model on the $N$ provided enrollment RIRs.
Several approaches for fine-tuning can be taken; recent work in LLM fine-tuning has shown the success of using low-rank adaptation, which adjusts a low-rank matrix of weights to adapt a model to a new domain with a limited amount of data \cite{hu2022lora}.
For a given $d \times d$ weight matrix $W$ of the model, two additional weight matrices $A,B \in \mathbb{R}^{d \times r}$ are introduced to update the weights as $\bar{W} = W + BA^T$, where $BA^T$ defines a rank $r$ weight matrix learned on the enrollment data.
We also experiment with fine-tuning the model by simply using the pre-trained model's weights to initialize a model trained on the provided training examples.
After fine-tuning, we perform inference by providing the coordinates of the target source and receiver locations for evaluation, along with the provided (or retrieved, for rooms 11--20) room mesh.

\section{Results}
We tested our method on the 5 provided examples from ``Room\_0'', with the results shown in Table~\ref{tab:results}, evaluating on the remaining provided examples in that room.
We can see in Table~\ref{tab:results} that our strategy's ability to reconstruct rooms is effective; we found that Rank-1 LoRA was the most effective method for capturing the EDF of the RIR.
We used the trained models for each room to generate data for Task 2 using the coordinates provided in the enrollment dataset, then fine-tuned the provided SDE model on this data and submitted the results.

\begin{table}[t!]
\centering
\label{tab:results}
\sisetup{
    reset-text-series = false, 
    text-series-to-math = true, 
    mode=text,
    tight-spacing=true,
    round-mode=places,
    round-precision=3,
    table-format=1.3,
    table-number-alignment=center
}
\caption{Performance on Room 0 of models using different pre-training sets and fine-tuning methods in terms of the error in RT60 [\%], EDF [dB], and DRR~[dB].}
\begin{tabular}{llSSS}
    \toprule
    Pre-training set& Fine-tuning method & {RT60}& {EDF} & {DRR}\\
    \midrule
    Retrieved-GWA & LoRA-1 & 0.090 & 0.520 & 3.009\\
    Retrieved-GWA & All Parameters & 0.1744 & 2.736 & 4.626\\
    Random-GWA & LoRA-1 & 0.186 & 0.652 & 5.155 \\
    Random-GWA & All Parameters & 0.792 & 0.562 & 11.172\\
    None & All Parameters & 0.417 & 4.416 & 8.917 	\\
    \bottomrule
\end{tabular}

\end{table}

\begingroup
\fontencoding{T1}\selectfont
\bibliographystyle{IEEEtran}
\bibliography{bibliography}
\endgroup
\end{document}